\documentclass[twocolumn, prl]{revtex4}
\pdfoutput=1
\usepackage{graphicx}
\usepackage{color}
\usepackage{epsfig}
\begin{document}
\title{ Signature of Quantum Criticality in the Density Profiles of Cold Atom Systems}

\author{Qi Zhou$^{+}$ and Tin-Lun Ho$^{\ast}$}
\affiliation{$^{+}$ Joint Quantum Institute and Condensed Matter Theory Center, University of Maryland, College Park, MD 20742, \\$^{\ast}$Department of Physics, The Ohio State University, Columbus, OH 43210}
\date{\today}

\begin{abstract} 
In recent years, there is considerable experimental
effort using cold atoms to study strongly correlated many-body
systems\cite{Bloch1, Bloch2, Esslinger,Chin,Ian,  Bloch3}. One class of phenomena of particularly interests 
is quantum critical (QC) phenomena. While prevalent in many materials, 
these phenomena are notoriously difficult theoretical problems due to the
vanishing of energy scales in QC region. So far, there are no
systematic ways to deduce QC behavior of bulk systems
from the data of trapped atomic gases. Here, we present a simple
algorithm to use the experimental density profile to determine the T=0 phase 
boundary of bulk systems, as well as the scaling functions in QC regime. We also present
another scheme for removing finite size effects of the trap. We
demonstrate the validity of our schemes using exactly
soluble models.
 \end{abstract}

\maketitle
In a recent paper, we gave a general scheme to derive thermodynamic functions of bulk systems at finite temperature from the density profile of trapped gases within local density approximation (LDA)\cite{HoZhou}. This scheme has recently been  implemented by Salomon's group at ENS to map out the global phase diagram and equation of state of  unitary Fermi gas\cite{Sal1, Sal2}.  The study of quantum phase transitions, however, involves  other challenges. First of all, these are phase transitions at $T=0$ caused by changes of the parameters of the system. In addition, there are systems with phase transitions only at $T=0$ but not  at finite temperatures. 
How to access the  {\em $T=0$} phase diagram  from the data of {\em trapped} gases at {\em non-zero temperature} is not obvious.  Secondly, cold atom experiments typically have $10^{5}$ to $10^{6}$ particles, which will give rise finite size effects. How to eliminate these effects systematically is also unclear.  The purpose here is to offer solutions to these problems. 

{\em (A) Algorithm for uncovering quantum criticality from non-uniform density profile:}
Let us consider the equation of state $n=n(T, \mu)$, where $n$ is the density and $\mu$ is  the chemical potential. 
Let   $\mu_{c}$ be the chemical potential at $T=0$ where the phase transition takes place. 
 Near  $\mu=\mu_{c}$, general arguments show that  when the dimensionality $d$  
 is below the upper critical dimension $d_c$,  $n=n(T, \mu)$ becomes 
\cite{QCn}
\begin{equation}
n(\mu, T) - n_{r}(T,\mu)= T^{\frac{d}{z}+1 -\frac{1}{\nu z}} {\cal G}\left(\frac{\mu-\mu_{c}}{T^{\frac{1}{\nu z}}}\right),
\label{n} \end{equation}
where $\nu$ is the  correlation length exponent, 
$n_{r}$ is the regular part of the density, and ${\cal G}(x)$ is a universal function describing the singular part of the density near criticality.  
(Above $d_c$,  $d/z+1 -1/ \nu z$ 
is replaced by the mean field value $1/z\nu$). 
Correspondingly, the compressibility $\kappa= \partial n/\partial \mu$ obeys  $T^{-1-\frac{d}{z} + \frac{2}{\nu z}} [\kappa(\mu, T)  -\kappa_{r}(\mu, T) ]= {\cal G}'\left(\frac{\mu-\mu_{c}}{
T^{ \frac{1}{\nu z}}}\right)$ for $d<d_{c}$.  For $d>d_{c}$,  $-1-\frac{d}{z} + \frac{2}{\nu z}$ is replaced by 0.  
 
Eq.(\ref{n}) is particularly useful when $n_{r}$ of one of the phases is known. For boson and femion  Mott insulators, $n_{r}$ are integers (hence $\kappa_{r}=0$).  For systems such as the fully spin polarized fermi gas, $n_{r}$ can be calculated analytically.  In such cases, if we plot  the ``scaled density" $A(\mu, T)\equiv T^{-1-d/z+1/\nu z}(n(\mu, T)-n_{r})$, or the scaled compressibility $C(\mu, T)\equiv T^{-1-\frac{d}{z} + \frac{2}{\nu z}} [\kappa(\mu, T)  -\kappa_{r}(\mu, T) ]$ 
 versus $\mu$ for different temperatures, then all curves will intersect at the same point $\mu=\mu_{c}$.   Once $\mu_{c}$ is determined, one can then plot
$A( \mu, T)$ (or $C(\mu, T)$) versus  $\tilde{u} \equiv (\mu - \mu_c)/T^{1/{\nu z}}$.   The scaled density curves for all temperatures will collapse into a single curve, which is the scaling function ${\cal G}(u)$ (or ${\cal G}'(u)$). 

This scheme can easily be implemented in cold atom experiments  within LDA, which takes the measured density $n^{ex}({\bf x})$ at point ${\bf x}$ at temperature $T$ as its bulk value $n(\mu, T)$ with $\mu$ replaced by $\mu({\bf x})= \mu - V({\bf x})$, where $V({\bf x})$ is the confining trap;  i.e. $n^{ex}({\bf x})= n(\mu({\bf x}), T)$. With $T$ and $\mu$ determined from the measured density profile\cite{HoZhou, HoZhou-temp}, $n=n(\mu, T)$ of the bulk system  is readily obtained from the non-uniform density profile $n^{ex}({\bf x})$ of the trapped gas.  
 To locate the critical point $\mu_c$, one simply plots  $A^{ex}({\bf r}) = T^{-1-d/z+ 1/\nu z} (n^{ex}({\bf r})-n_{r})$ 
 versus $\mu({\bf r})$ for different samples at different temperatures $T$ and looks for their  common intersection. 
 Such intersection will only occurs if $z$ and $\nu$ assume their correct values.  Recalling that  $z$ is typically either 1 or 2, and $\nu$ can be estimated from our knowledge of various universality classes,  the number of trials needed in practice to 
 produce the common intersection is very small. 
 
Even if $n_{r}$ is unknown, its regularity allows one to expand it around $(\mu=\mu_{c}, T=0)$,  $n_{r}(T,\mu)= n_{r}(0,\mu_{c}) + \sum_{(n,m)\neq(0,0)} \alpha_{n,m}T^{n}(\mu-\mu_{c})^{m}$. One can then implement the previous scheme by retaining the first few terms in the series, treating their coefficients $\alpha_{n,m}$ as parameters to be adjusted to obtain a crossing of the scaled densities or compressibilities at various temperatures.  

{\em (B) Quantum criticality of 1D hard core bosons:} 
To demonstrate our scheme, we use 1D hard core bosons and 1D free Fermi gas as examples. Since both systems have exact solutions in confining traps and in free space, we can then characterize the accuracy of LDA by comparing the density calculated from it with that calculated from the solution in a harmonic trap, (denoted as $n^{ex}(x)$), which now plays the role of experimental data.   Our hamiltonian is 
$H = -J\sum_{\langle {\bf R, R'}\rangle} (b^{\dagger}_{{\bf R}} b^{}_{\bf R'} + h.c.) + V_{\bf R} n_{\bf R}$,  
where ${\bf R}$  labels the lattice sites , $\langle {\bf R,R'} \rangle$ are nearest neighbors, $b_{\bf R}^{\dagger}$ creates a boson at  ${\bf R}$, $V_{\bf R}= \frac{1}{2}M\omega^2 {\bf R}^2$ is the harmonic potential, and $n_{\bf R}= b^{\dagger}_{\bf R} b_{\bf R}^{} $.  The hardcore constraint is implemented by restricting $n_{\bf R}=0$ or 1. 
It is well known that the wavefunction for 1D hardcore bosons, $\Psi_{B}(R_1, R_2, ..., R_{N})$, is that of a free Fermi gas  $\Psi_{F}(R_1, R_2, ..., R_{N})$ multiplied by a factor $A= \prod_{R>R'}\epsilon(R - R')$, where $\epsilon(x) =1$ or $(-1)$ if $x>0$ ($x<0$); i.e. $\Psi_{B} = A \Psi_{F} $.  
The density profile of hard core bosons is then given by that of the free Fermi gas, \begin{equation}
n^{ex}(x) = \sum_{n=0,1,2, ..} |u_{n}(x)|^2 f(E_{n}), 
\label{1Dnex} \end{equation}
where $u_{n}$ is the eigenfunction of $H$ with energy $E_{n}$, and $f(x)= 1/(e^{(x-\mu)/T}+1)$ is the Fermi function. For homogeneous systems (where $V_{\bf R}=0$), the density is 
\begin{equation}
n(\mu, T) =  \int^{\pi/a}_{-\pi/a} \frac{ {\rm d}k}{2\pi} f(\epsilon_{k})
= \int^{2J}_{-2J}  {\rm d}E \mathcal{N}(E) f(E)
\label{1Dn} \end{equation}
where $\epsilon_{k} = -2J{\rm cos}ka$, 
$\mathcal{N}(E)= (\pi a)^{-1} (4J^2 - E^2)^{-1/2}$ is the density of states, and $a$ is the lattice constant. 
Note that both Eq.(\ref{1Dnex}) and (\ref{1Dn}) also apply to free fermions.

 \begin{figure}[tbp]
\begin{center}
\includegraphics[width=3in]{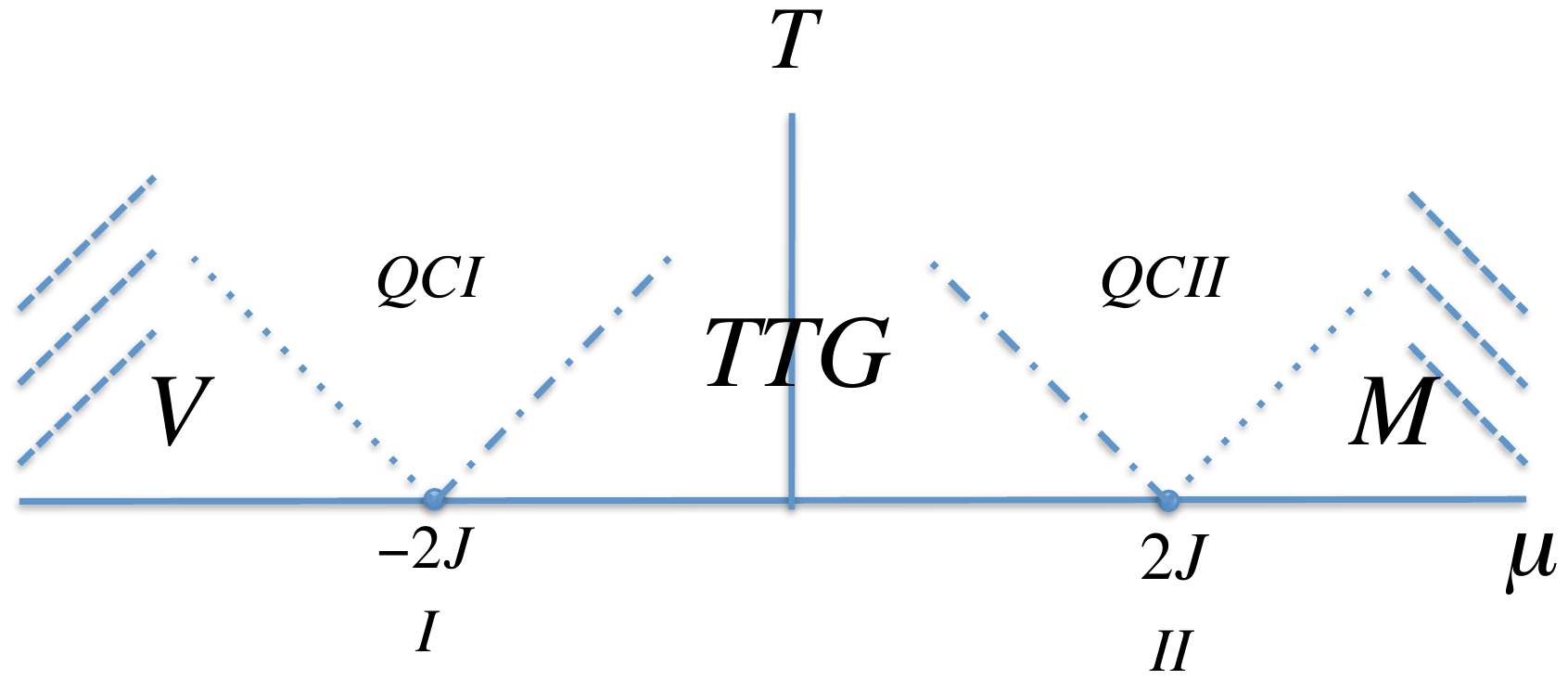}
\end{center}
\caption{Phase diagram of 1D hard core Bose gas. $I$ and $II$ are the quantum critical points of vacuum (V) to Tomonaga-Luttinger Liquid (TLL), and Mott (M)  to TLL transitions. The quantum critical regions associated with quantum critical point $I$ and $II$ are denoted as $QCI$ and $QCII$, given by 
$|\mu+2J|< T$ and $|\mu-2J|<T$ respectively. The shaded region in the vacuum and the Mott phase represent the low fugacity regime of particle and holes.}
 \end{figure}
The $T=0$ phase diagram of this system is shown in Figure 1. There are two QC points,  $\mu_{c} =-2J$ and $+2J$ in the $T-\mu$ plane, corresponding to the transition from vacuum ($n=0)$ to a Tomonaga-Luttinger Liquid  (TLL)  and from the TLL to Mott phase $(n=1)$; denoted as $I$ and $II$, respectively.  The region $T>|2J+\mu|$ ($QCI$ in Fig.1) is a quantum critical region, where thermal excitations dominate over the excitation energy of the system (measured from $\mu$).  Similar region appears near QC point $II$ by particle-hole symmetry. 
 Near $QCI$, $\mu_{c} = -2J$, and the density and compressibility scale as 
$ \lambda n(\mu, T) = Li_{1/2}(-e^{(\mu+2J)/T})$
and
$\lambda T \frac{dn}{d\mu} = Li_{-1/2}( -e^{(\mu+2J)/T})$, 
where $\lambda= h/\sqrt{2\pi m^{\ast} T}$, $m^{\ast}= \hbar^2/(Ja^2)$ is the effective mass, and 
$Li_{n}(x) = \sum_{k=0, 1, 2..} x^{k}/k^{n}$ is the polylog function.
 \begin{figure}[tbp]
\begin{center}
\includegraphics[width=3in]{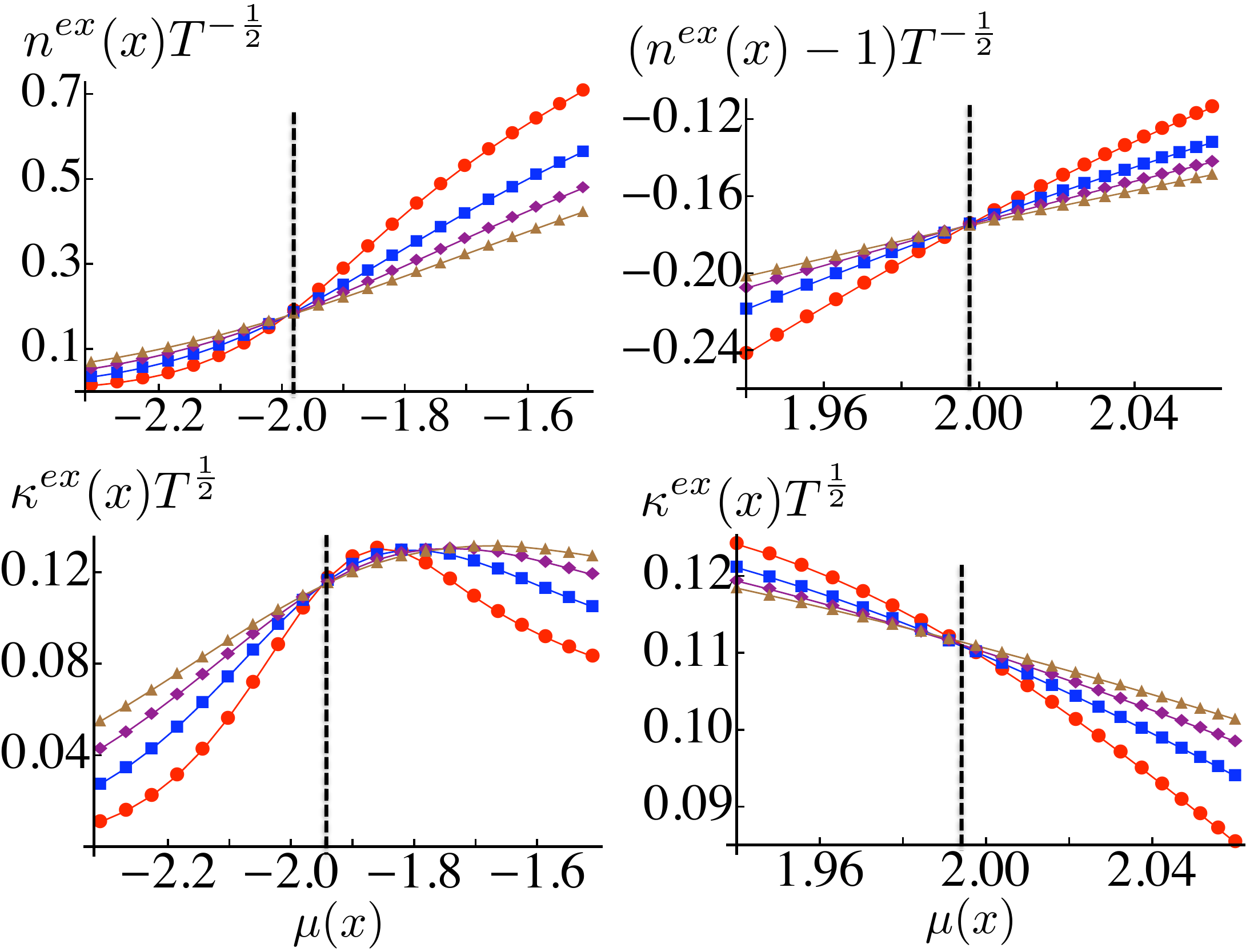}
\end{center}
\caption{ Scaled density (top) and scaled compressibility (bottom) vs. $\mu(x) = \mu-\frac{1}{2}M \omega^2 x^2$ at different temperatures $T/J = 0.1, 0.15, 0.2, 0.25$ (red, blue, purple, and brown curves), for the 1D hardcore Bose gas discussed in the text.  Left and right panels are results for transitions $I$ and $II$, respectively. When $z$ is chosen to be $2$ and $\nu=1/2$, all scaled density(compressibility) intersect at $\mu^{\ast}_{I}/J=-1.98(-1.94)$ and $\mu^{\ast}_{II}/J= 1.998(1.991)$. The exact values $\mu_{c}$  of these critical points are $-1$ and $1$.}
 \end{figure}

  \begin{figure}[tbp]
\begin{center}
\includegraphics[width=3in]{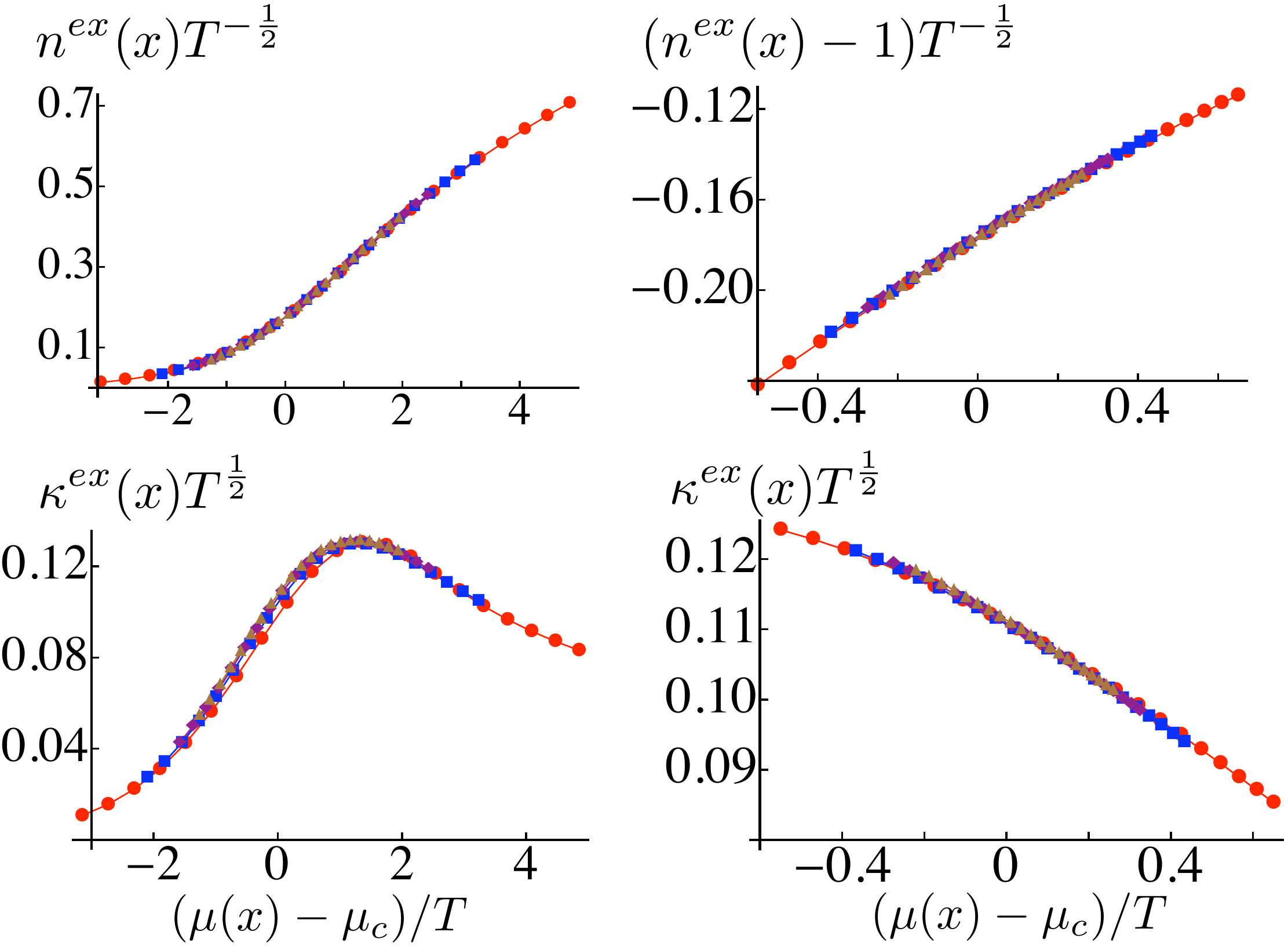}
\end{center}
\caption{Scaled density (top) and scaled compressibility (bottom)  vs. $(\mu(x)-\mu_c)/T^{z\nu}$. All the data collapse onto a single curve. The parameters are the same as Fig. 2.}
 \end{figure}

 {\em (C)  Demonstration of algorithms in (A):}
For explicit demonstration, we 
consider a hard core Bose gas with $N= 265$ particles in a trap of frequency $\omega$ adjusted to produce a Mott phase at the trap center, corresponding to $M\omega^2 a^2/(2J)=0.0001$.  Choosing $z=2$, and $\nu=1/2$ (which are the correct values for this model), the scaled quantities are  $A^{ex}(x) = n^{ex}(x)T^{-1/2}$ and $C^{ex}(x)= \kappa^{ex}(x) T^{1/2}$.  
In Figure 2, we have plotted $A^{ex}(x)$ and $C^{ex}(x)$ versus $\mu(x)$ for different temperatures. We find that for $T\lesssim 0.3J$,  all the curves at different temperatures intersect at $\mu^{\ast}_{I}/J=-1.98$ and $\mu^{\ast}_{II}/J= 1.998$, corresponding to the critical points  $I$ and $II$ in Fig.1. Comparing with the exact value $\mu_{c}/J = -2$ and $2$, one sees that  LDA is accurate to $99\%$ even for a system with $10^2$ particles. 
At $T>0.3J$,  the scaled compressibility at different temperatures fail to intersect at the same point, showing that QC properties  appear only when $T\lesssim 0.3J$.  Similar intersections are found when we plot the scaled compressibility with local chemical potential (see Figure 2). To construct the scaling function, we plot the scaled density and scaled compressibility at different $T$  against the scaled variable $\mu(x)/T^{1/(z\nu)}$ around QC$I$ and $II$.  We see from Figure 3 that all data collapse onto a single curve near $\mu^{\ast}$, which is the scaling function for these QC points.  

   \begin{figure}[tbp]
\begin{center}
\includegraphics[width=3in]{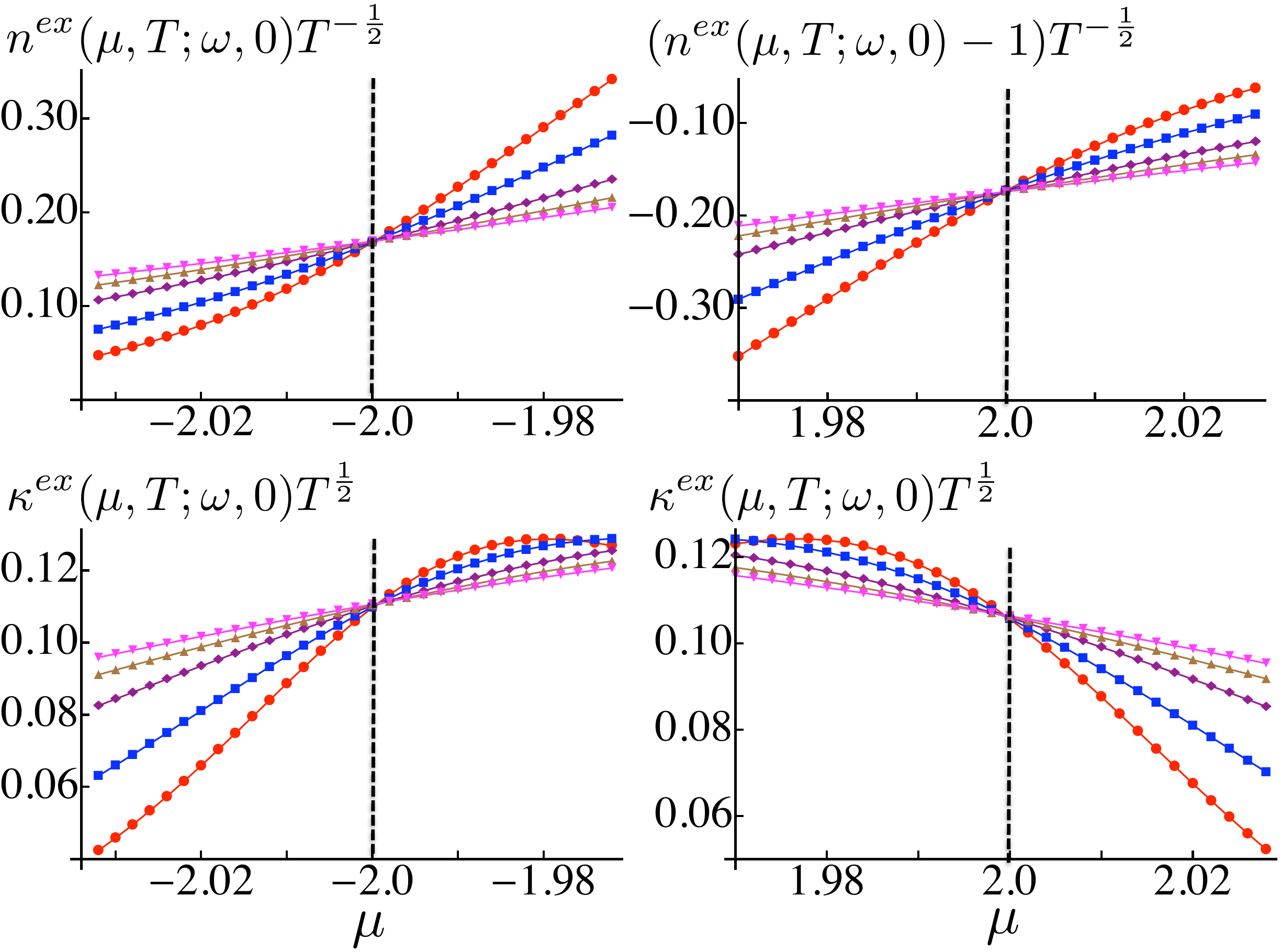}
\end{center}
\caption{ Determining $\mu_{c}$ of bulk systems using a finite-size scaling scheme:  the scaled density and scaled compressibility are plotted against $\mu$ for different $T/J=0.02, 0.03, 0.05, 0.07$  (red, blue purple, brown, and pink curves) for 1D hard core bosons in harmonic traps. The value ${\cal D}$ defined in the text is set equal to 0.25. }
 \end{figure}

{\em (D) Finite size scaling}:  
Although the small error of the LDA scheme (about $1\%$)  is hardly of  concern to current experiments, it is a matter of principle whether it can be eliminated. This question is not academic,  for it will also be of experimental relevance as  the resolution of density measurements continue to improve. Moreover, it is also relevant for understanding the results of numerical calculations.  Here, we show that this error can be corrected by a (more involved) algorithm that make use of finite size scaling rather than LDA. 
Using renormalization group arguments\cite{scaling1,scaling2}, it has been shown that for a bulk system close to a QC point, if one switches on a harmonic potential, then  the  singular part of a thermodynamic quantity
near ${\bf x}=0$  follows certain scaling relations.  The validity of these scaling relations has been well tested for quantum spin systems\cite{scaling1,scaling2}.  For example, under a scale change $b$, the singular part of the density below $d_c$ scales as 
\begin{equation}
n_s(\mu, T,  \omega^2, {\bf x})=b^{-(d+z)+1/\nu} g(\bar{\mu} b^{\frac{1}{\nu}}, Tb^z,\omega^2 b^{y}, {\bf x}/b),
\label{ns} \end{equation}
where $\overline{\mu}=\mu-\mu_{c}$, $y=2+1/\nu$, and $g$ is  scaling function. 
Choosing $Tb^z=1$, the scaled density  $A(\mu, T; D, {\bf x})$
$\equiv$ $ T^{-1-d/z +1/\nu z}n_s(T,\mu, \omega^2, {\bf x})$ at point ${\bf x}$  
satisfies 
\begin{equation}
A(\mu, T,  \omega^2, {\bf x}) =g\left(\bar{\mu}/T^{\frac{1}{\nu z}}, \omega^{2}/T^{\frac{y}{z}}, {\bf x}T^{\frac{1}{z}}\right).
\label{finite} \end{equation}
Defining  $D\equiv \omega^{2}/T^{\frac{y}{z}}$, Eq.(\ref{finite}) becomes
\begin{equation}
\overline{A}(\mu, T | D, {\bf x}) =g\left(\bar{\mu}/T^{\frac{1}{\nu z}}, D, {\bf x}T^{\frac{1}{z}}\right),
\label{finite2} \end{equation}
where  $\overline{A}(\mu, T|D, {\bf x}) \equiv A(\mu, T, D T^{\frac{y}{z}}, {\bf x}) $.
We then have $\overline{A}(\mu, T |D, {\bf 0}) =g\left(\bar{\mu}/T^{\frac{1}{\nu z}}, D, {\bf 0}\right)$. Hence, 
if we plot $\overline{A}$ versus $\mu$ at ${\bf x =0}$ for different $T$ with $D$ held fixed,
 then different curves with different $T$ (but same $D$) will intersect at $\mu_{c}$ since  $\overline{A}(\mu_{c}, T|D, {\bf 0})=g\left(0, D,{\bf 0}\right)$. A similar analysis shows scaled compressibility $ T^{-1-d/z +2/\nu z}\kappa_s(T,\mu, \omega^2, {\bf x})$ at constant $D$ also intersect at $\mu_c$.  

  \begin{figure}[tbp]
\begin{center}
\includegraphics[width=3in]{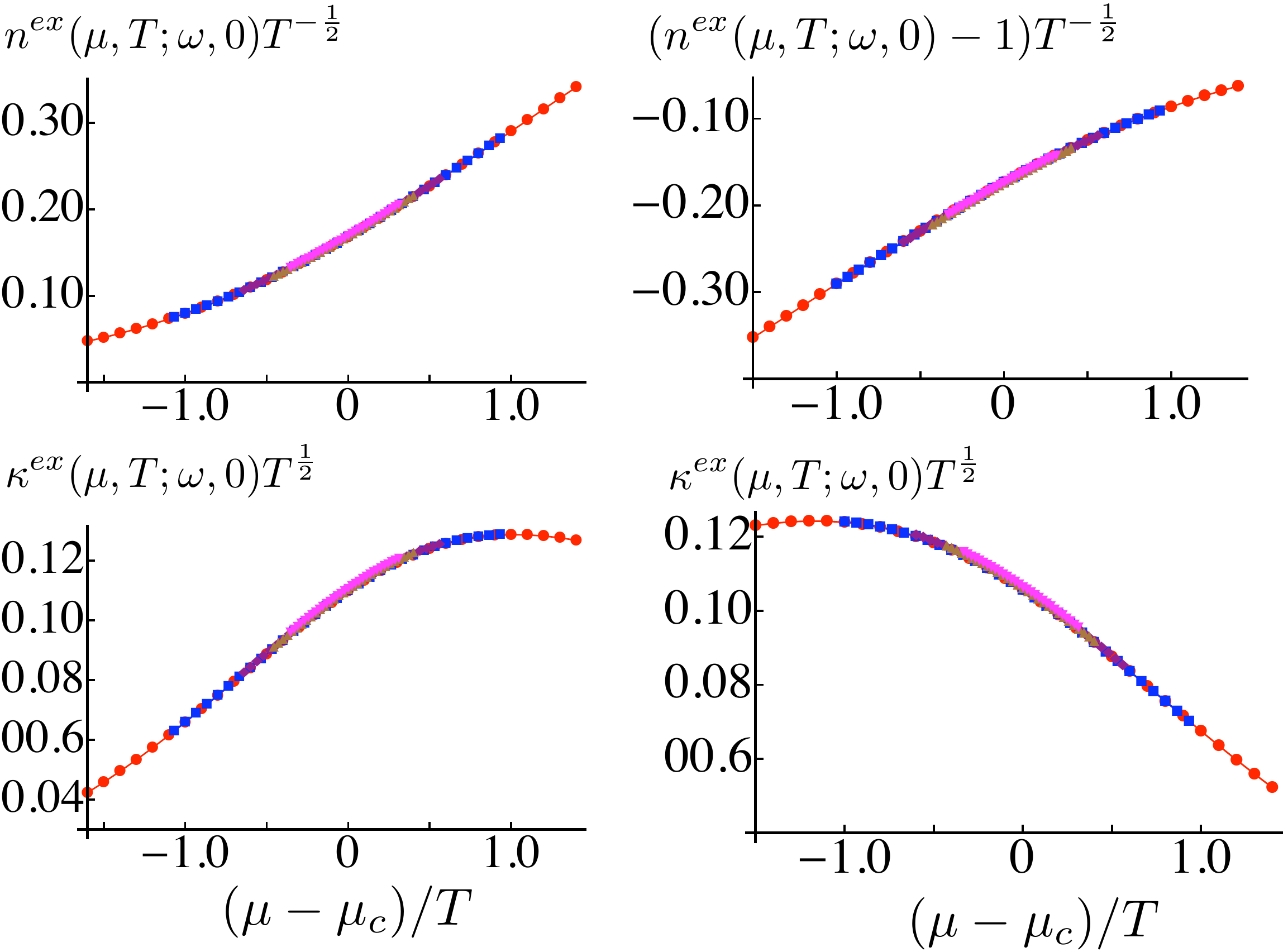}
\end{center}
\caption{ Scaled density (top) and scaled compressibility (bottom)  vs. $(\mu-\mu_c)/T^{1/{z\nu}}$ with $\nu z=1$. All curves in Figure 4 collapse onto a single curve. The parameters are the same as Fig. 4. }
 \end{figure}

 For 1D hard core bosons, $z=2$, $\nu=1/2$, $n_s=n$, we have $\overline{A}(\mu, T|D, x)= n^{ex}(\mu, \omega, T; x) T^{-1/2}$ near $x=0$. 
Our scheme is to plot  $ n^{ex}(\mu, \omega, T; 0) T^{-1/2}$ versus $\mu$ for a family of samples with different 
$T$ and $\omega$ but with identical values for the ratio $D= \omega^2/T^{\frac{y}{z}}$. These plots, together with similar plots for the compressibility,  are given in Figure 4,  with the ratio ${\cal D}\equiv  (\frac{1}{2}M\omega^2 a^2/J)/(T/J)^{y/z} = 0.25$.  We find that, to machine accuracy, different curves intersect exactly at $\mu_{c}/J=2.0$. Moreover, if we plot $n^{ex}(\mu, T; \omega, 0)T^{-1/2}$ against the variable 
$\tilde{u} = (\mu-\mu_{c})/T^{1/z\nu}$, we see in Fig. 5 that all curves collapse to the same curve, which is the scaling function $g(u, D, 0)$.

By holding $D$ fixed, we need to know the densities
 $\{ n^{ex}(\mu, T; \omega)\} $ for  a range of $T$, $\mu$ and $\omega$ values. 
 Thus, this scheme is clearly more laborious than the LDA scheme. 
 Still, it is of conceptual importance to demonstrate that the $\mu_{c}$ of bulk systems can be obtained exactly without errors.  The recent work by Salomon's group~\cite{Sal2} mapping out the equation of state of a unitary gas indicates that the construction of the entire family $\{ n^{ex}(\mu, T; \omega)\} $ is feasible. Our scheme will also be useful in numerical studies of trapped atoms, especially for unsolved problems where the existence of quantum criticality is in question.

  {\em (E) Finite temperature phase transition:} Previously, we \cite{HoZhou} have pointed out that within LDA, 
 the phase boundary $\mu_{c}(T)$ of a  finite temperature continuous  phase transition will show up as a kink in the 
  in the compressibility $\kappa^{ex}(x)= \partial n^{ex}(x)/\partial \mu(x)$ at position $x^{\ast}$ such that  $\mu_{c}(T) = \mu(x^{\ast})$.
As with the quantum critical case discussed in main text, LDA will contain a systematic error due to finite size effects, even though it is very small.  This error can also be corrected in a similar  way.  The analog of Eq.~(\ref{ns}) for a finite temperature phase transition is\cite{Campostrini} 
   \begin{equation}  
  n_{s}(T, \mu, \omega^2, {\bf x}) = b^{-d+1/\nu}n_{s}(\overline{\mu}(T)b^{1/\nu}, \omega^{2}b^{y}, {\bf x}/b),
\label{Tscaling}  \end{equation}
where $\overline{\mu}(T)\equiv \mu - \mu_{c}(T)$ and $\nu$ is the correlation length exponent for finite $T$ transition. 
(For example,  $\nu=0.67$ for the 3D xy universality class.) 
Setting $\omega^2 b^{y}=1$, Eq.~(\ref{Tscaling}) implies the scaling form 
\begin{equation}
  n_{s}(T, \mu, \omega^2, {\bf x}) =\omega^{x}{\cal F}(\omega^{-2/y\nu} \overline{\mu}(T), \omega^{2/y} {\bf x}),
\end{equation}
where $x= (2/y)(d-1/\nu)$ and ${\cal F}$ is a scaling function.   Similar analysis gives $\kappa_{s}(T, \mu, \omega^2, {\bf x}) =\omega^{(2/y)(d-2/\nu)}{\cal F}(\omega^{-2/y\nu} \overline{\mu}(T), \omega^{2/y} {\bf x})
$.  The full density $n$, however,  contains both singular ($n_{s}$) and regular ($n_{r}$)  contributions;  i.e. 
   $n= n_{r}+n_{s}$. The unknown regular contribution $n_{r}$ can be eliminated by measuring the difference in density profiles at different trap frequencies $\omega\neq \omega_1$ relative to some reference value $\omega_{1}$.  Fixing $\omega_1$, the function
\begin{equation}
B_{T}(\mu; \omega) =\frac{ n(T, \mu, \omega_{1}^2, {\bf 0})- n(T, \mu, \omega^2, {\bf 0})}{\omega^{x}_{1}-\omega^{x}}
\end{equation}
will assume the value ${\cal F}(0, {\bf 0})$ at $\mu=\mu_{c}(T)$ for all $\omega$. This means that if we plot $B_{T}(\mu; \omega)$ as a function of $\mu$, then different curves for different $\omega$ will intersect at the exact phase boundary $\mu_{c}(T)$.

The fact that the entire phenomenology of quantum criticality of {\em bulk} systems -- its existence, its dynamical critical exponent, its scaling function, and the location of the quantum critical point -- can all be deduced from the $T\neq 0$ density profile of trapped gases as demonstrated here further adds to the list of valuable information deducible from density measurements~\cite{HoZhou}. 
At present, the field of quantum gases is moving rapidly in the direction of high precision measurements, a direction essential for realizing the full power of quantum simulation~\cite{HoZhou}. The extraordinarily high resolution for density imaging recently achieved by Greiner's group~\cite{Ott, Greiner} shows that the accuracy needed for deducing bulk properties from trapped gases has been achieved. The recent success of Salomon's group in deducing the phase diagram of a unitary gas~\cite{Sal1,Sal2} demonstrates the feasibility of our algorithms. All these developments strongly suggest the power of quantum simulation is ready to be harvested. 

This work was supported by two grants from the Army Research Office with funding from the DARPA OLE program, and by the NSF grant DMR0907366 awarded to TLH. QZ is supported by  JQI-NSF-PFC and ARO-DARPA-OLE.

\end{document}